\documentclass[english,aps]{revtex4-1}
\usepackage[T1]{fontenc}
\usepackage[latin9]{inputenc}
\usepackage{color}
\usepackage{array}
\usepackage{mathrsfs}
\usepackage{amsmath}
\usepackage{amssymb}
\usepackage{graphicx}

\makeatletter

\providecommand{\tabularnewline}{\\}

\usepackage{babel}

\makeatother

\usepackage{babel}
\begin{document}
\title{Incorporating the Coulomb potential into a finite, unitary perturbation
theory}
\author{Scott E. Hoffmann}
\address{School of Mathematics and Physics,~~\\
 The University of Queensland,~~\\
 Brisbane, QLD 4072~~\\
 Australia}
\email{scott.hoffmann@uqconnect.edu.au}

\begin{abstract}
We have constructed a perturbation theory to treat interactions that
can include the Coulomb interaction, describing a physical problem
that is often encountered in nuclear physics. The Coulomb part is
not treated perturbatively; the exact solutions are employed. The
method is an extension of the results presented in Hoffmann (2021
\textit{J. Math. Phys.} 62 032105). It is designed to calculate phase
shifts directly rather than the full form of the wavefunctions in
position space. We present formulas that allow calculation of the
phase shifts to second order in the perturbation. The phase shift
results to second order, for a short-range potential, were compared
with the exact solution, where we found an error of third order in
the coupling strength. A different model, meant as a simple approximation
of nuclear scattering of a proton on Helium-4 and including a Coulomb
potential and a spherical well, was constructed to test the theory.
The wavepacket scattering formalism of Hoffmann (2017 \textit{J. Phy.
B: At. Mol. Opt. Phys} 50 215302), known to give everywhere finite
results, was employed. We found physically acceptable results and
a cross section of the correct order of magnitude.
\end{abstract}
\maketitle

\section{Introduction}

A fundamental problem in nuclear physics is to treat the scattering
from a system with a Coulomb interaction and a short-range nuclear
interaction. This is known as the Coulomb-nuclear interference problem
(\citet{Deltuva2005,Durand2020,Franco1973,Islam1967,Petrov2018,West1968}).
Here we treat this problem within nonrelativistic quantum mechanics
but note that the velocities involved are often relativistic when
describing the results of experiments. 

Our treatment involves partial wave analysis. This was thought to
be not applicable to the Coulomb scattering problem, as the sum over
the angular momentum quantum number, $l,$ diverges in a plane wave
treatment. In \citet{Hoffmann2017a}, treating the scattering of a
wavepacket by a Coulomb potential was found to introduce a convergence
factor into this sum and to lead to physical results, thereby making
the method applicable to the Coulomb case.

In that paper, probabilities of wavepacket to wavepacket transitions
were calculated and found to be everywhere finite and less than or
equal to unity (including in the forward scattering direction). A
simple formula from that paper relates these probabilities to differential
cross sections.

It is well known that the solutions of the pure Coulomb problem are
not accessible by perturbation theory. While the first-order contribution
can be made finite in some perturbation theories, all higher-order
corrections in the pure Coulomb case diverge (\citet{Dalitz1951}).
In our unitary perturbation method, we do not find finiteness even
for the first-order contribution if the Coulomb potential is treated
perturbatively.

In a recent paper (\citet{Hoffmann2021a}), this author introduced
a unitary perturbation theory for the radial Schrödinger equation
(to treat spherically symmetric potentials). The calculations were
done in momentum space rather than position space, with resulting
simplifications. The goal of the method is to calculate the phase
shifts for the scattering problem. Along with the wavepacket parameters,
these contain all the information necessary to describe a scattering
process. This paper presents formulas for their calculation up to
second order in the coupling strength (which has the magnitude of
the velocity in the denominator. See eq. (\ref{eq:2.16.5})). Only
the case of $s$-wave scattering ($l=0$) was treated there. The first
aim of this paper (in section \ref{sec:Unitary-perturbation-method})
is to extend the results to general nonnegative integers, $l,$ and
to test the results by comparison with an exactly solvable model.
Mathematical methods needed for the derivation are given in appendix
A.

The method relates the free momentum eigenvectors to the interacting
eigenvectors of the same momentum by a unitary transformation. The
exponential generator of this transformation is written as a power
series in a dimensionless coupling constant. The transformation of
the Hamiltonian from the free case to the interacting case (free plus
potential) gives equations for the terms in the generator to each
order in the coupling constant. These can be solved, taking care to
eliminate divergences by imposing a rule of principal part integration.
The unitarity of the transformation guarantees that the interacting
state vector remains conveniently normalized at each order in the
approximation.

In section \ref{sec:Comparison-with-the} we choose an exactly solvable
model and compare our results with the exact solutions for the phase
shifts.

In section \ref{sec:Perturbation-around-exact}, as the central result
of this paper, we propose to use the known exact solutions for the
Coulomb potential and perturb around them with a suitably well-behaved
nuclear potential. The mathematical methods needed for this procedure
are given in this section and in appendix B. The total phase shift
is a sum of the Coulomb phase shift, $\sigma_{l}(p),$ and a correction,
$\delta_{l}^{+}(p).$ We present formulas for the calculation of the
$\delta_{l}^{+}(p)$ up to second order in the nuclear coupling strength,
$\eta^{+}$ (eq. (\ref{eq:4.14})), and to all orders in the Coulomb
coupling strength, $\eta^{C}$ (eq. (\ref{eq:4.6})).

For a non-Coulomb potential, the phase shifts are calculated directly
using formulas that are in terms of matrix elements of the potential
between free spherical waves. In the case where we treat a potential
that is the sum of a Coulomb potential and a perturbation, the matrix
elements are between solutions of the radial Schrödinger equation
with a Coulomb potential. In principle, there is no restriction on
the strength of the Coulomb interaction.

Then, in section \ref{sec:Example:-Spherical-well}, we choose a spherical
well as the ``nuclear'' part of the potential, with parameters chosen
to model the scattering of a proton and a $\phantom{\vert}_{2}^{4}\mathrm{He}$
nucleus, and calculate differential cross sections. The main purpose
of this exercise is to demonstrate that we obtain everywhere finite
results that are physically realistic.

This perturbation theory will generally be applicable for sufficiently
high momenta (as $\eta^{+}$ is the expansion parameter), so that
this method could be improved by extending it to the relativistic
regime. It is the intention of the author to do that in a future paper.

Conclusions follow in section \ref{sec:Conclusions}.

This method is to be contrasted with other methods for calculating
phase shifts. Several approaches (\citet{Vigo-Aguiar2005,Simos1997})
involve numerically integrating a full solution in the direction of
the radial coordinate, $r,$ then extracting the phase shift from
the asymptotic behaviour for large $r.$ These methods have the advantage
that they do not rely on a perturbative system; they obtain a numerical
approximation to the full solution. The method presented here was
derived using only the asymptotic behaviour of the solutions and does
not calculate the wavefunction at lower $r.$

Hence these two approaches can be seen as complementary. For a system
that is judged to be perturbative, the method presented here allows
fast, direct determination of the phase shift contributions at first
and second order in the coupling. For nonperturbative systems, a numerical
integration method would be appropriate.

The WKB method (\citet{Messiah1961,Kramers1929}) was used by Bethe
on the Coulomb-nuclear problem to derive the phase that bears his
name (\citet{Bethe1958,Islam1967}). This is the relative phase between
the Coulomb and nuclear terms in the total scattering amplitude for
this problem. If we applied this approximation method to the regime
where the energy is larger that the maximum of the potential (no tunneling),
best results would be obtained for large energies, much greater than
this maximum. Hence the conditions for an accurate approximation are
similar to those for other perturbation theories.

Throughout this paper, we use Heaviside-Lorentz units, in which $\hbar=c=\epsilon_{0}=\mu_{0}=1.$

\section{\label{sec:Unitary-perturbation-method}Unitary perturbation method
to second order}

Our goal is to solve the radial Schrödinger equation
\begin{equation}
\{-\frac{1}{2m}\frac{d^{2}}{dr^{2}}+\frac{l(l+1)}{2mr^{2}}+V(r)\}y_{l}^{(i)}(r,p)=\frac{p^{2}}{2m}y_{l}^{(i)}(r,p)\label{eq:2.0.1}
\end{equation}
for the interacting wavefunctions
\begin{equation}
y_{l}^{(i)}(r,p)=\langle\,r\,|\,p,l;i\,\rangle\label{eq:2.0.2}
\end{equation}
and an interaction potential $V.$ The boundary condition is
\begin{equation}
y_{l}^{(i)}(r,p)\sim(pr)^{1+\epsilon}\quad\mathrm{with}\quad\epsilon\geq0,\label{eq:2.0.3}
\end{equation}
to ensure regularity of the three-dimensional solution
\begin{equation}
\psi_{l}(\boldsymbol{r})=\frac{y_{l}^{(i)}(r,p)}{r}Y_{lm}(\hat{\boldsymbol{r}})\label{eq:2.0.4}
\end{equation}
at the origin.

The central proposal leading to this unitary perturbation theory is
that the energy eigenvectors and operators of the free and interacting
systems, at equal momenta, can be related by a unitary transformation.
If the state vectors transform as
\begin{equation}
|\,p,l;i\,\rangle=U_{if}|\,p,l;f\,\rangle,\label{eq:2.1}
\end{equation}
then the Hamiltonians must be related by
\begin{equation}
U_{if}H_{f}U_{if}^{\dagger}=H_{i}=H_{f}+\lambda U,\label{eq:2.2}
\end{equation}
so that we have the eigenvalue relation
\begin{equation}
H_{i}|\,p,l;i\,\rangle=(H_{f}+\lambda U)|\,p,l;i\,\rangle=U_{if}H_{f}U_{if}^{\dagger}U_{if}|\,p,l;f\,\rangle=\frac{p^{2}}{2m}|\,p,l;i\,\rangle.\label{eq:2.3}
\end{equation}
We have written the potential as $V=\lambda U$ to show explicitly
the dependence on a dimensionless coupling constant, $\lambda.$

\textcolor{red}{Note that $i$ refers here to the free or unperturbed
system. Then $f$ refers to the perturbed system, with Hamiltonian
given by the first part of Eq. \ref{eq:2.3}.}

We write
\begin{equation}
U_{if}=e^{-i\Theta},\label{eq:2.4}
\end{equation}
in terms of a generator, $\Theta.$

We assume that this generator can be expressed as a power series in
the coupling constant
\begin{equation}
\Theta=\lambda\Theta^{(1)}+\frac{1}{2}\lambda^{2}\Theta^{(2)}+\dots,\label{eq:2.5}
\end{equation}
to second order. Note $\Theta$ must vanish at $\lambda=0$ to give
$U_{if}=1$ there.

\textcolor{red}{The author is grateful to a referee, who pointed out
that this is known as the Magnus expansion. Useful results are contained
in the references \citep{Blanes2009,Blanes2010,Casas2007}.}

We expand $U_{if}$ to $\mathcal{O}(\lambda^{2})$
\begin{equation}
U_{if}\cong1-i\lambda\Theta^{(1)}-\frac{i}{2}\lambda^{2}\Theta^{(2)}-\frac{1}{2}\lambda^{2}\Theta^{(1)2}\label{eq:2.6}
\end{equation}
and insert this into eq. \ref{eq:2.2}, then equate like powers of
$\lambda.$ This gives
\begin{align}
[H_{f},\lambda\Theta^{(1)}] & =-iV,\nonumber \\{}
[H_{f},\lambda^{2}\Theta^{(2)}] & =[\lambda\Theta^{(1)},V].\label{eq:2.7}
\end{align}

Solving these for the free matrix elements gives
\begin{equation}
\langle\,k_{1},l;f\,|\,\lambda\Theta^{(1)}\,|\,k_{2},l;f\,\rangle=-i\frac{2mV_{l}(k_{1},k_{2})}{k_{1}^{2}-k_{2}^{2}}\label{eq:2.8}
\end{equation}
and
\begin{multline}
\langle\,k_{1},l;f\,|\,\lambda^{2}\Theta^{(2)}\,|\,k_{2},l;f\,\rangle=\\
\quad\frac{-i}{k_{1}^{2}-k_{2}^{2}}P\int_{0}^{\infty}dk^{\prime}\,\{\frac{2mV_{l}(k_{1},k^{\prime})2mV_{l}(k^{\prime},k_{2})}{k_{1}^{2}-k^{\prime2}}-\frac{2mV_{l}(k_{1},k^{\prime})2mV_{l}(k^{\prime},k_{2})}{k^{\prime2}-k_{2}^{2}}\}.\label{eq:2.9}
\end{multline}
Consequently
\begin{equation}
\langle\,k_{1},l;f\,|\,\lambda^{2}\Theta^{(1)2}\,|\,k_{2},l;f\,\rangle=-P\int_{0}^{\infty}dk^{\prime}\,\frac{2mV_{l}(k_{1},k^{\prime})}{k_{1}^{2}-k^{\prime2}}\frac{2mV_{l}(k^{\prime},k_{2})}{k^{\prime2}-k_{2}^{2}}.\label{eq:2.10}
\end{equation}

Here the free momentum matrix elements of the potential are
\begin{equation}
V_{l}(k_{1},k_{2})=\int_{0}^{\infty}dr\,y_{l}^{(f)}(r,k_{1})V(r)y_{l}^{(f)}(r,k_{2}),\label{eq:2.11}
\end{equation}
with
\begin{equation}
y_{l}^{(f)}(r,k)=\sqrt{\frac{2}{\pi}}\,krj_{l}(kr)\label{eq:2.11.1}
\end{equation}
in terms of spherical Bessel functions (\citet{Abramowitz1972}).
As $r\rightarrow\infty,$ these have the asymptotic form
\begin{equation}
y_{l}^{(f)}(r,k)\rightarrow\sqrt{\frac{2}{\pi}}\,\sin(kr-l\frac{\pi}{2}).\label{eq:2.11.2}
\end{equation}
So for the matrix element integrals to converge as $r\rightarrow\infty$
requires that $V(r)$ fall off faster than $1/r,$ excluding the Coulomb
potential. Near $r=0,$ the strongest bound comes from examining the
radial Schrödinger equation in that region. If $V(r)\sim r^{\nu}$
near the origin, we must have $\nu>-2.$

We deal with the singularities in our method by imposing the rule
that principal part integration must be used. We will see that this
leads to finite results in agreement with the exact solution for a
model potential (in section \ref{sec:Comparison-with-the}). Here
$P\int$ indicates a principal part integral, defined in our case
as
\begin{equation}
P\int_{-1}^{1}dx\,\frac{f(x)}{x}=\lim_{\epsilon\rightarrow0^{+}}\{\int_{-1}^{-\epsilon}dx\,\frac{f(x)}{x}+\int_{\epsilon}^{1}dx\,\frac{f(x)}{x}\}.\label{eq:2.12}
\end{equation}

The perturbed wavefunction to second order is then
\begin{multline}
y_{l}^{(2)}(r,p)=\langle\,r;f\,|\,1-i\lambda\Theta^{(1)}-\frac{i}{2}\lambda^{2}\Theta^{(2)}-\frac{1}{2}\lambda^{2}\Theta^{(1)2}\,|\,p,l;f\,\rangle\\
=y_{l}^{(f)}(r,p)-P\int_{0}^{\infty}dk\,y_{l}^{(f)}(r,k)\frac{2mV_{l}(k,p)}{k^{2}-p^{2}}\\
+\frac{1}{2}P\int_{0}^{\infty}dk\,y_{l}^{(f)}(r,k)\,P\int_{0}^{\infty}dk^{\prime}\,\frac{2mV_{l}(k,k^{\prime})}{k^{2}-k^{\prime2}}\frac{2mV_{l}(k^{\prime},p)}{k^{\prime2}-p^{2}}\\
\quad-\frac{1}{2}P\int_{0}^{\infty}dk\,\frac{y_{l}^{(f)}(r,k)}{k^{2}-p^{2}}\,P\int_{0}^{\infty}dk^{\prime}\,\{\frac{2mV_{l}(k,k^{\prime})2mV_{l}(k^{\prime},p)}{k^{2}-k^{\prime2}}-\frac{2mV_{l}(k,k^{\prime})2mV_{l}(k^{\prime},p)}{k^{\prime2}-p^{2}}\}.\label{eq:2.13}
\end{multline}
Note that this is real, as will be the case at all orders.

To extract the phase shifts, we only need the asymptotic form of this
wavefunction as $pr\rightarrow\infty.$ It is expected to take the
form
\begin{align}
y_{l}^{(2)}(r,p) & \rightarrow\sqrt{\frac{2}{\pi}}\,\sin(pr-l\frac{\pi}{2}+\delta_{l}^{(1)}(p)+\delta_{l}^{(2)}(p))\nonumber \\
 & \cong\{1-\frac{1}{2}\delta_{l}^{(1)2}(p)\}\sqrt{\frac{2}{\pi}}\,\sin(pr-l\frac{\pi}{2})+\{\delta_{l}^{(1)}(p)+\delta_{l}^{(2)}(p)\}\sqrt{\frac{2}{\pi}}\,\cos(pr-l\frac{\pi}{2}),\label{eq:2.14}
\end{align}
where $\delta^{(1)}(\delta^{(2)})$ is the first (second) order contribution
to the phase shift. In this way we can identify the contributions
to the phase shift from the form of our asymptotic wavefunction.

Using the mathematical methods of Appendix A, we found
\begin{align}
y_{l}^{(2)}(r,p) & \rightarrow\{1-\frac{1}{2}\eta^{2}v_{l}(p,p)^{2}\}\sqrt{\frac{2}{\pi}}\,\sin(pr-l\frac{\pi}{2})\nonumber \\
 & \quad+\{-\eta\,v_{l}(p,p)+\eta^{2}(\Delta_{-}(l,p)+\Delta_{+}(l,p)+\Delta_{\infty}(l,p))\}\sqrt{\frac{2}{\pi}}\,\cos(pr-l\frac{\pi}{2}),\label{eq:2.15}
\end{align}
where
\begin{align}
\Delta_{-}(l,p) & =\frac{2}{\pi}\int_{-1}^{1}dx\,\frac{\{v_{l}(p,p(1+x))^{2}-v_{l}(p,p(1-x))^{2}\}}{x(4-x^{2})},\nonumber \\
\Delta_{+}(l,p) & =-\frac{1}{\pi}\int_{-1}^{1}dx\,\frac{\{v_{l}(p,p(1+x))^{2}+v_{l}(p,p(1-x))^{2}\}}{4-x^{2}},\nonumber \\
\Delta_{\infty}(l,p) & =\frac{2}{\pi}\int_{2}^{\infty}dz\,\frac{v_{l}(p,pz)^{2}}{z^{2}-1},\label{eq:2.16}
\end{align}
and
\begin{equation}
\eta=\frac{\lambda}{p/m}\quad\mathrm{and}\quad v_{l}(k_{1}k_{2})=\frac{\pi}{\lambda}V_{l}(k_{1},k_{2}).\label{eq:2.16.5}
\end{equation}
Thus our predictions for the phase shift contributions are
\begin{align}
\delta_{l}^{(1)}(p) & =-\eta\,v_{l}(p,p),\nonumber \\
\delta_{l}^{(2)}(p) & =\eta^{2}(\Delta_{-}(l,p)+\Delta_{+}(l,p)+\Delta_{\infty}(l,p)).\label{eq:2.17}
\end{align}
The equation for the first order contribution is well-known (\citet{Messiah1961},
their eq. X.74).

All of these integrals converge for the model system we will consider
in section \ref{sec:Comparison-with-the}. More generally, for the
class of well-behaved potentials we are considering, we find $v_{l}(p,0)=0$
and that $v_{l}(p,pz)$ will be an asymptotically decreasing function
of $z.$ So $\Delta_{\pm}(l,p)$ will always converge. Since there
is a factor of $1/(z^{2}-1)$ in the integrand for $\Delta_{\infty}(l,p),$
that integral will also always converge.

\section{\label{sec:Comparison-with-the}Comparison with the exact solutions
for the spherical barrier/well}

To have an exact solution with which to compare our results, we consider
the radial Schrödinger equation with potential
\begin{equation}
V^{\mathrm{b/w}}(r)=\begin{cases}
\frac{\lambda}{R} & 0<r<R,\\
0 & r>R,
\end{cases}\label{eq:3.1}
\end{equation}
a spherical barrier $(\lambda>0)$ or well $(\lambda<0)$ with height
$V_{0}=\lambda/R.$ This was done in \citet{Hoffmann2021a}, but only
for $l=0.$

The full solutions of energy $p^{2}/2m$ are proportional to $y_{l}^{(f)}(\kappa^{\prime}\frac{r}{R})$
on the inner region (to satisfy the boundary condition of vanishing
at least as fast as $Cr$ at $r=0$) and are linear combinations of
$y_{l}^{(f)}(\kappa\frac{r}{R})$ and $\tilde{y}_{l}^{(f)}(\kappa\frac{r}{R})$
on the outer region, where
\begin{equation}
\tilde{y}_{l}^{(f)}(z)=\sqrt{\frac{2}{\pi}}\,z\,n_{l}(z)\label{eq:3.2}
\end{equation}
and the $n_{l}(z)$ are the spherical Neumann functions, which diverge
at the origin (\citet{Abramowitz1972}, their section 10.1.3). Here
\begin{equation}
\kappa=pR\quad\mathrm{and}\quad\kappa^{\prime}=\sqrt{\kappa^{2}-2\eta(p)\kappa}\label{eq:3.3}
\end{equation}
with $\eta(p)=\lambda/(p/m).$

We merely quote the result for the phase shifts
\begin{equation}
\delta_{l}^{\mathrm{exact}}(p)=\mathrm{Arg}(A_{l}(p)-iB_{l}(p)),\label{eq:3.4}
\end{equation}
with
\begin{align}
A_{l}(p) & =\kappa^{2}j_{l}(\kappa^{\prime})n_{l}^{\prime}(\kappa)-\kappa^{\prime}\kappa j_{l}^{\prime}(\kappa^{\prime})n_{l}(\kappa),\nonumber \\
B_{l}(p) & =\kappa^{\prime}\kappa j_{l}^{\prime}(\kappa^{\prime})j_{l}(\kappa)-\kappa^{2}j_{l}(\kappa^{\prime})j_{l}^{\prime}(\kappa).\label{eq:3.5}
\end{align}

Expansion in $\eta$ gives the first-order contribution
\begin{equation}
\delta_{0}^{(1)\mathrm{exact}}=-\eta\,(1-\mathrm{sinc}(2\kappa)).\label{eq:3.5.5}
\end{equation}

We use the results of section \ref{sec:Unitary-perturbation-method}
with 
\begin{multline}
V_{l}^{\mathrm{b/w}}(p,p(1+x))=\frac{2\lambda\kappa^{2}}{\pi}(1+x)\int_{0}^{1}d\rho\,\rho^{2}j_{l}(\kappa\rho)j_{l}(\kappa(1+x)\rho)\\
\quad=\frac{2\lambda\kappa}{\pi}\frac{1+x}{2+x}\{\frac{j_{l-1}(\kappa)j_{l}(\kappa(1+x))-j_{l-1}(\kappa(1+x))j_{l}(\kappa)}{x}\\
-j_{l-1}(\kappa(1+x))j_{l}(\kappa)\},\label{eq:3.6}
\end{multline}
free from singularities on $x\in[-1,\infty)$ for $l\geq1.$ For $l=0$
we use $j_{-1}(z)=-n_{0}(z)$ (\citet{Abramowitz1972}, their eq.
(10.1.12)).

\begin{figure}
\begin{centering}
\includegraphics[width=17cm]{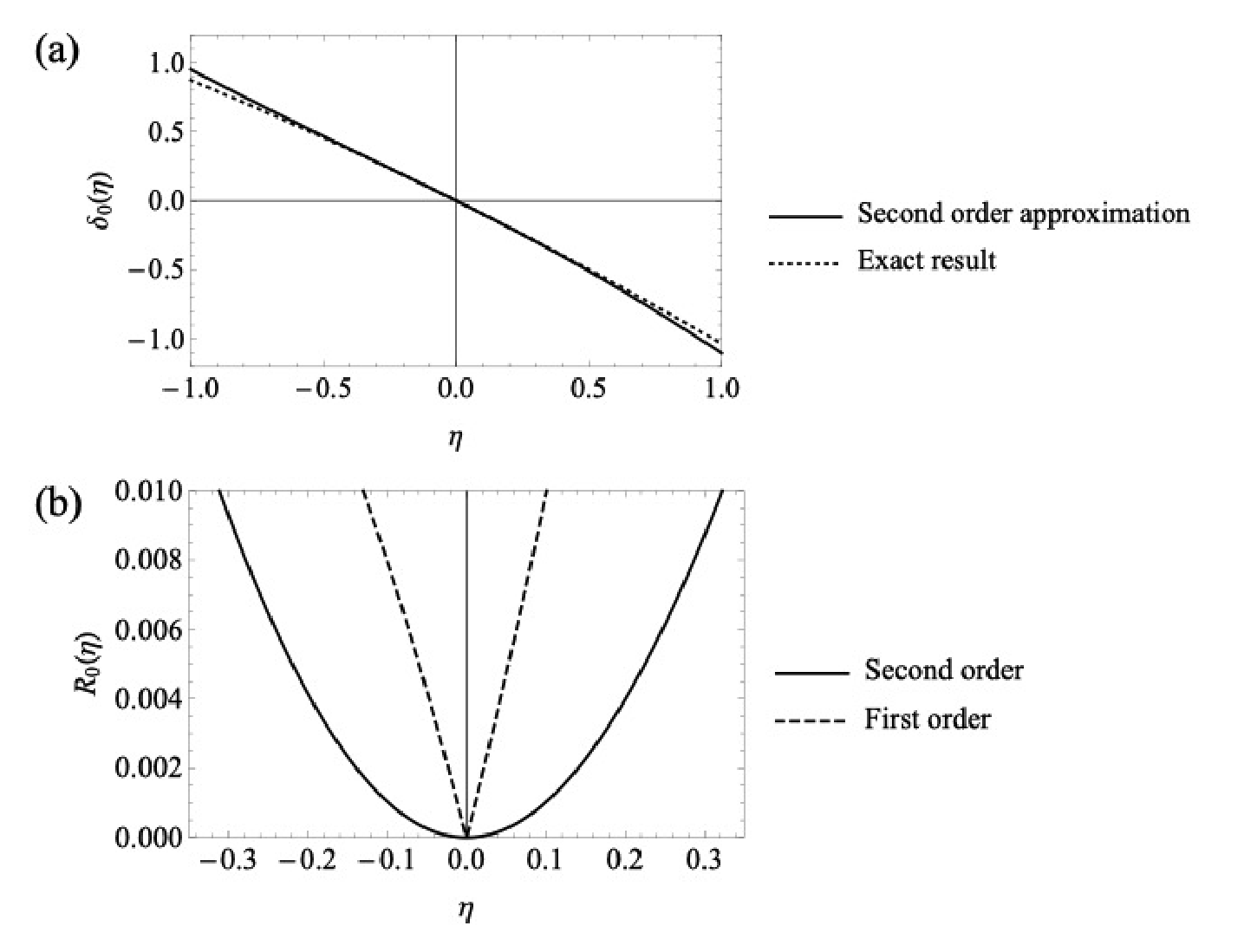}
\par\end{centering}
\caption{\label{fig:(a)-Second-order}(a) Total phase shift to second order
for $l=0,$ $\kappa=10,$ (b) relative error in the result.}

\end{figure}

For $l=0,$ $\kappa=10,$ we find the results in figure \ref{fig:(a)-Second-order}(a).
Here
\begin{equation}
R_{l}(\eta)=\left|\frac{\delta_{l}(\eta)-\delta_{l}^{\mathrm{exact}}(\eta)}{\delta_{l}^{\mathrm{exact}}(\eta)}\right|\label{eq:3.7}
\end{equation}
is the relative error in the approximation. In figure \ref{fig:(a)-Second-order}(b)
we show the relative error for the second order approximation and
the first order approximation, noting that the former gives a significantly
better approximation than the latter.

\begin{figure}
\begin{centering}
\includegraphics[width=17cm]{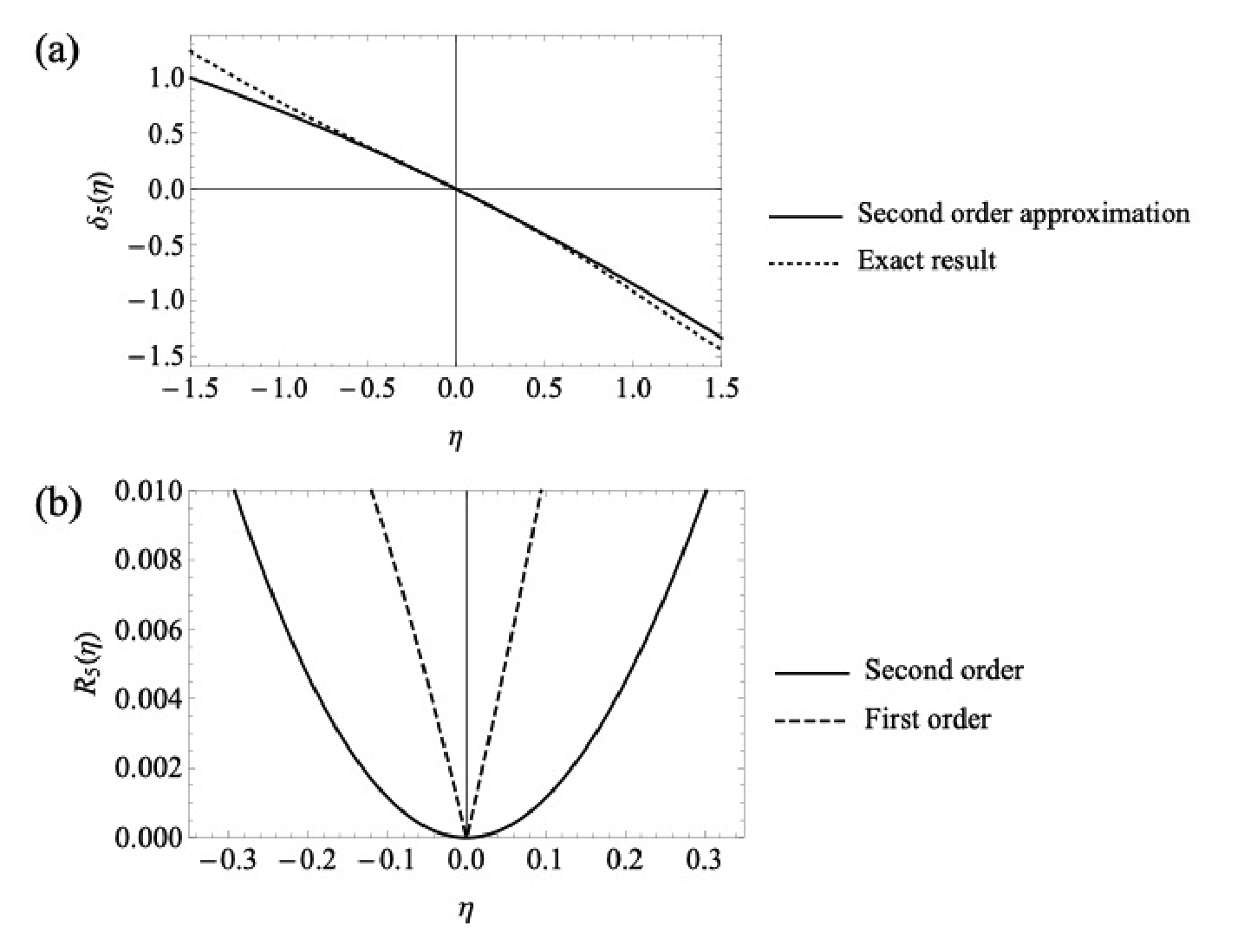}
\par\end{centering}
\caption{\label{fig:Results-for-}(a) Total phase shift to second order for
$l=5,$ $\kappa=10,$ (b) relative error in the result.}

\end{figure}

For $l=5,$ $\kappa=10,$ we find a similar region to the $l=0$ case
over which the approximation is useful, shown in figure \ref{fig:Results-for-}.

As is typical for short-range potentials, the phase shifts fall off
rapidly with $l,$ so that, in practice, only a small number of $l$
values need be used for a good approximation of the cross section.

\section{\label{sec:Perturbation-around-exact}Perturbation around exact Coulomb
solutions}

As discussed in the introduction, it is well known that perturbation
theory, in any of its forms, applied to the Coulomb potential,
\begin{equation}
V^{(C)}(r)=\frac{Z_{t}Z_{p}\alpha}{r},\label{eq:4.1}
\end{equation}
produces divergent contributions at second and higher order in $\alpha$.
Here $Z_{t}$ is the atomic number of the target, $Z_{p}$ is the
atomic number of the projectile and $\alpha\cong1/137$ is the fine
structure constant.

For the unitary perturbation theory presented here, not even the first
order contribution is finite. The $s$-wave ($l=0$) matrix elements
of the potential are
\begin{equation}
V_{0}^{(C)}(k_{1},k_{2})=\int_{0}^{\infty}dr\,\sqrt{\frac{2}{\pi}}\,\sin(k_{1}r)\frac{Z_{t}Z_{p}\alpha}{r}\sqrt{\frac{2}{\pi}}\,\sin(k_{1}r)=\frac{Z_{t}Z_{p}\alpha}{2\pi}\ln\left(\frac{(k_{1}+k_{2})^{2}}{(k_{1}-k_{2})^{2}}\right).\label{eq:4.2}
\end{equation}
Clearly $V_{0}^{(C)}(p,p)$ diverges for all $p.$

However, the exact solutions of the Coulomb problem,
\begin{equation}
\{-\frac{1}{2m}\frac{d^{2}}{dr^{2}}+\frac{l(l+1)}{2mr^{2}}+\frac{Z_{t}Z_{p}\alpha}{r}\}y_{l}^{(C)}(r,p)=\frac{p^{2}}{2m}y_{l}^{(C)}(r,p),\label{eq:4.3}
\end{equation}
are known (\citet{Messiah1961}). They are real and given by
\begin{equation}
y_{l}^{(C)}(r,p)=\sqrt{\frac{2}{\pi}}\,c_{l}(\eta^{C}(p))\,e^{ipr}\,(pr)^{l+1}\,F(l+1+i\eta^{C}(p),2l+2;-2ipr),\label{eq:4.4}
\end{equation}
in terms of a degenerate hypergeometric function, $F,$ (\citet{Gradsteyn1980},
their section 9.21. They use the notation $F\rightarrow\Phi.$). The
coefficients are
\begin{equation}
c_{0}=\left(\frac{2\pi\eta}{e^{2\pi\eta}-1}\right)^{\frac{1}{2}}\label{eq:4.4.1}
\end{equation}
and, for $l\geq1,$
\begin{equation}
c_{l}=\frac{c_{0}}{(2l+1)!!}\prod_{s=1}^{l}\left(1+\frac{\eta^{2}}{s^{2}}\right)^{\frac{1}{2}}.\label{eq:4.4.2}
\end{equation}
These solutions are orthonormalized to
\begin{equation}
\int_{0}^{\infty}dr\,y_{l}^{(C)}(r,k_{1})y_{l}^{(C)}(r,k_{2})=\delta(k_{1}-k_{2}).\label{eq:4.5}
\end{equation}
Here
\begin{equation}
\eta^{C}(p)=\frac{Z_{t}Z_{p}\alpha}{p/m}\label{eq:4.6}
\end{equation}
is a dimensionless measure of the coupling strength. These solutions
have the asymptotic behaviour, for $pr\rightarrow\infty,$
\begin{equation}
y_{l}^{(C)}(r,p)\rightarrow\sqrt{\frac{2}{\pi}}\,\sin(pr-\eta^{C}(p)\ln(2pr)-l\frac{\pi}{2}+\sigma_{l}(p)),\label{eq:4.7}
\end{equation}
where the Coulomb phase shifts, $\sigma_{l}(p),$ are
\begin{equation}
\sigma_{l}(p)=\mathrm{Arg}(\Gamma(l+1+i\eta^{C}(p))),\label{eq:4.8}
\end{equation}
(\citet{Messiah1961}).

We propose perturbing around these solutions for a potential
\begin{equation}
V(r)=\frac{Z_{t}Z_{p}\alpha}{r}+V^{+}(r)\label{eq:4.9}
\end{equation}
that includes the Coulomb potential and a perturbation, $V^{+}(r).$
This situation is commonly encountered in nuclear scattering of charged
particles, where they interact through the Coulomb potential and also
a short range nuclear contribution. We note that $V^{+}(r)$ must
be in the class for which this perturbation theory is applicable,
which will be the same class as for perturbation around free spherical
waves, discussed in section \ref{sec:Unitary-perturbation-method}.
In particular, $V^{+}(r)$ must vanish faster than $1/r$ as $r\rightarrow\infty.$

So we take
\begin{equation}
H_{0}=-\frac{1}{2m}\frac{d^{2}}{dr^{2}}+\frac{l(l+1)}{2mr^{2}}+\frac{Z_{t}Z_{p}\alpha}{r}\label{eq:4.10}
\end{equation}
as the unperturbed Hamiltonian and develop the unitary perturbation
theory for this problem, with full Hamiltonian
\begin{equation}
H=H_{0}+V^{+}.\label{eq:4.10.5}
\end{equation}

Results are written in terms of the potential matrix elements
\begin{equation}
\mathscr{\mathcal{V}}_{l}(k_{1},k_{2})=\frac{\pi}{\lambda^{+}}\int_{0}^{\infty}dr\,y_{l}^{(C)}(r,k_{1})V^{+}(r)y_{l}^{(C)}(r,k_{2}),\label{eq:4.11}
\end{equation}
which will always converge for the abovementioned constraints on $V^{+}.$

In Appendix B, we show that the presence of the logarithmic phase,
$-\eta(p)\ln(2pr),$ and the Coulomb phase shifts, $\sigma_{l}(p),$
in the asymptotic form given in eq. (\ref{eq:4.7}) does not prevent
us from obtaining results very similar to those of eq. (\ref{eq:2.17}).
The asymptotic forms of the solutions to second order in $\eta^{+}$
take the forms
\begin{equation}
y_{l}^{(C)}(r,p)\rightarrow\sqrt{\frac{2}{\pi}}\,\sin(pr-\eta^{C}(p)\ln(2pr)-l\frac{\pi}{2}+\sigma_{l}(p)+\delta_{l}^{(+)}(p)),\label{eq:4.11.5}
\end{equation}
with
\begin{equation}
\delta^{(+)}(p)\cong\delta_{l}^{(+,1)}(p)+\delta_{l}^{(+,2)}(p)\label{eq:4.11.6}
\end{equation}
and
\begin{align}
\delta_{l}^{(+,1)}(p) & =-\eta^{+}\mathcal{V}_{l}(p,p),\nonumber \\
\delta_{l}^{(+,2)}(p) & =\eta^{+2}(\Delta_{-}^{(+)}(l,p)+\Delta_{+}^{(+)}(l,p)+\Delta_{\infty}^{(+)}(l,p)),\label{eq:4.12}
\end{align}
with
\begin{align}
\Delta_{-}^{(+)}(l,p) & =\frac{2}{\pi}\int_{-1}^{1}dx\,\frac{\{\mathcal{V}_{l}(p,p(1+x))^{2}-\mathcal{V}_{l}(p,p(1-x))^{2}\}}{x(4-x^{2})},\nonumber \\
\Delta_{+}^{(+)}(l,p) & =-\frac{1}{\pi}\int_{-1}^{1}dx\,\frac{\{\mathcal{V}_{l}(p,p(1+x))^{2}+\mathcal{V}_{l}(p,p(1-x))^{2}\}}{4-x^{2}},\nonumber \\
\Delta_{\infty}^{(+)}(l,p) & =\frac{2}{\pi}\int_{2}^{\infty}dz\,\frac{\mathcal{V}_{l}(p,pz)^{2}}{z^{2}-1}.\label{eq:4.13}
\end{align}
Here
\begin{equation}
\eta^{+}=\frac{\lambda^{+}}{p/m}.\label{eq:4.14}
\end{equation}

Again, all of these integrals will always converge for potentials
in the restricted class. The solutions are real, as will be the case
at all orders.

\section{\label{sec:Example:-Spherical-well}Example: Spherical well nuclear
potential}

A simple model used in nuclear physics takes the internuclear potential
to be the sum of a Coulomb potential and a spherical well of the form
considered in section \ref{sec:Comparison-with-the} . Our aim was
to construct a realistic description of a proton incident on a nuclear
target. The first choice was $\eta^{+}=-1,$ a significant interaction
strength at the limits of applicability of our perturbation theory.
Nuclear potential well depths have been measured for a large number
of isotopes using slow neutron scattering (\citet{Czachor2011}).
From that reference, Helium-4 ($Z_{t}=2$) was selected as the target,
with a potential height of $V_{0}=-30.2\,\mathrm{MeV}$. Incident
on the target is a proton of momentum $p=237\,\mathrm{MeV}$ ($E=29.5\,\mathrm{MeV}$).

From the reduction of the scattering of two particles to that of a
single projectile on a fixed target, the mass must be replaced by
the reduced mass in the expressions for $\eta^{+}$ and $\eta^{C}:$
\begin{equation}
m\rightarrow\mu=\frac{m_{\mathrm{He}}m_{p}}{m_{\mathrm{He}}+m_{p}}=749\,\mathrm{MeV}.\label{eq:5.1}
\end{equation}
Then the relevant quantity to determine the velocity is $p/\mu.$
This gives a velocity of $\beta=v/c=0.302.$ Then the gamma factor
is $\gamma=1.05.$ One limitation of the model is that this velocity
enters slightly into the relativistic regime. Absent are relativistic
corrections, which would be at the $5\,\%$ level. The other main
limitation is modelling the nuclear potential as a spherical well.

An estimate of the nuclear radius comes from \citet{Czachor2011}
\begin{equation}
R=1.3\,\mathrm{fm}\,A^{\frac{1}{3}}=2.06\,\mathrm{fm},\label{eq:5.2}
\end{equation}
where $A=4$ is the mass number. For consistency with the equations
\begin{equation}
\kappa=pR,\quad V_{0}=\frac{\lambda^{+}}{R},\quad\eta^{+}=\frac{\lambda^{+}}{p/\mu}\quad\mathrm{and}\quad\eta^{C}=\frac{Z_{t}Z_{p}\alpha}{p/\mu},\label{eq:5.3}
\end{equation}
we find
\begin{equation}
\kappa=2.48,\quad\lambda^{+}=-0.316,\quad\eta^{C}=0.0461.\label{eq:5.4}
\end{equation}

Our numerical calculations give the phase shifts at first and second
order, and the total, in table \ref{tab:First--and-second-order}.
We compare these to the exact phase shifts for the spherical well
only, with these model parameters.

\begin{table}
\begin{centering}
\begin{tabular}{|>{\centering}m{1.5cm}|>{\centering}m{2cm}|>{\centering}m{2cm}|>{\centering}m{2cm}|>{\centering}m{2cm}|}
\hline 
$l$ & $\delta_{l}^{(+,1)}$ & $\delta_{l}^{(+,2)}$ & $\delta_{l}^{(+,1)}+\delta_{l}^{(+,2)}$ & $\delta_{l}^{\mathrm{exact}}$ (Nuclear)\tabularnewline
\hline 
$\ 0\ $ & 1.230 & -0.316 & 0.914 & 0.805\tabularnewline
\hline 
$\ 1\ $ & 0.651 & 0.299 & 0.950 & 0.906\tabularnewline
\hline 
$\ 2\ $ & 0.136 & 0.050 & 0.186 & 0.232\tabularnewline
\hline 
$\ 3\ $ & 0.015 & 0.003 & 0.018 & 0.020\tabularnewline
\hline 
$\ 4\ $ & 0.001 & 0.000 & 0.001 & 0.001\tabularnewline
\hline 
\end{tabular}
\par\end{centering}
\caption{\label{tab:First--and-second-order}First- and second-order phase
shift contributions and their sum for the model parameters.}
\end{table}

In \citet{Hoffmann2017a}, a formalism was developed to describe the
scattering of a Gaussian wavepacket with a momentum width, $\sigma_{p},$
small compared to the incoming average momentum, $p,$ incident on
a fixed target Coulomb potential. We calculated wavepacket-to-wavepacket
transition probabilities, but a simple formula presented there allows
the calculation of cross sections. The fact that a probability can
never rise greater than unity was responsible for the observation
that the Coulomb probability takes a finite value in the forward direction.
Calculation of plane wave scattering predicts a divergence in the
forward direction.

It is a simple matter to adapt this formalism to the model considered
here. This treatment, with a Coulomb potential included, is expected
to avoid a singularity in the forward direction. We chose the momentum
width relative to the incident average momentum such that
\begin{equation}
\epsilon=\frac{\sigma_{p}}{p}=0.001.\label{eq:5.5}
\end{equation}

The formula we used for the differential cross section (including
the behaviour around the forward direction), modified from \citet{Hoffmann2017a}
to include the nuclear phase shifts, is
\begin{equation}
\frac{d\sigma^{+}}{d\Omega}=\frac{1}{4p^{2}}\left|\sum_{l=0}^{\infty}(2l+1)e^{-2\epsilon^{2}(l+\frac{1}{2})^{2}}e^{i2\sigma_{l}(p)}e^{i2(\delta_{l}^{(+,1)}(p)+\delta_{l}^{(+,2)}(p))}P_{l}(\cos\theta)\right|^{2}.\label{eq:5.6}
\end{equation}
(We ignored time delays/advancements.) This has been tested to give
a cross section in very close agreement with the Rutherford cross
section (for $\delta_{l}^{(+,1)}(p)+\delta_{l}^{(+,2)}(p)=0$), except
close to $\theta=0,$ where a finite result is predicted.

A numerical problem arose. If this sum was evaluated as written, the
resulting cross section appeared to include only the Coulomb part.
It was necessary to use the identity
\begin{equation}
e^{i2\delta_{l}}=1+2i\,e^{i\delta_{l}}\,\sin(\delta_{l}),\label{eq:5.7}
\end{equation}
with
\begin{equation}
\delta_{l}=\delta_{l}^{(+,1)}(p)+\delta_{l}^{(+,2)}(p)\label{eq:5.8}
\end{equation}
to separate the sum into two parts. Adding the resulting sums and
taking the modulus-squared gave a differential cross section with
both Coulomb and nuclear contributions.

\begin{figure}
\begin{centering}
\includegraphics[width=8cm]{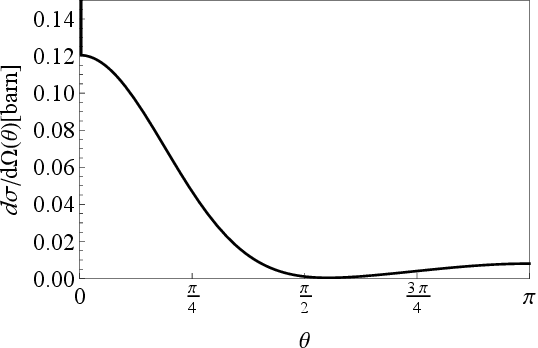}
\par\end{centering}
\caption{\label{fig:Differential-cross-section-1}Differential cross section
for spherical well only: exact solution with model parameters.}

\end{figure}

First we plot (in fig. \ref{fig:Differential-cross-section-1}) the
differential cross section for the nuclear spherical well only, using
the exact phase shifts with our model parameters, for comparison.
We used
\begin{equation}
\frac{d\sigma_{\mathrm{s}}^{N}}{d\Omega}=\frac{1}{p^{2}}\left|\sum_{l=0}^{l_{\mathrm{max}}}(2l+1)\,e^{i\delta_{l}^{\mathrm{exact}}}\sin(\delta_{l}^{\mathrm{exact}})\,P_{l}(\cos\theta)\right|^{2},\label{eq:5.9}
\end{equation}
with $l_{\mathrm{max}}=4.$ Note $\exp(-2\epsilon^{2}(l+\frac{1}{2})^{2})=1+\mathcal{O}(\epsilon^{2}).$
This formula also comes from using the separation in eq. (\ref{eq:5.7})
and leaves out the strong, narrow contribution around the forward
direction, although we plotted it in figure \ref{fig:Differential-cross-section-1}.
The wavepacket formalism adds nothing here except smearing over the
angular scale $\Delta\theta\sim0.001$ and that narrow peak around
$\theta=0$, not a delta function but a function of the form
\begin{equation}
\frac{d\sigma_{\mathrm{forward}}^{N}}{d\Omega}=\frac{1}{16p^{2}}\frac{1}{\epsilon^{4}}\,e^{-\theta^{2}/4\epsilon^{2}}.\label{eq:5.10}
\end{equation}

We display the cross sections in barns ($1\,\mathrm{barn}=10^{-28}\,\mathrm{m}^{2}=100\,\mathrm{fm^{2}}$)
and the momentum in MeV, including the strong, narrow contributions
around the forward direction. If
\[
\frac{d\sigma}{d\Omega}=\frac{1}{p^{2}}f
\]
in Heaviside-Lorentz units, the conversion is given by
\[
\frac{d\sigma}{d\Omega}[\mathrm{barn}]=\frac{389}{p[\mathrm{MeV}]^{2}}f.
\]

\begin{figure}
\begin{centering}
\includegraphics[width=17cm]{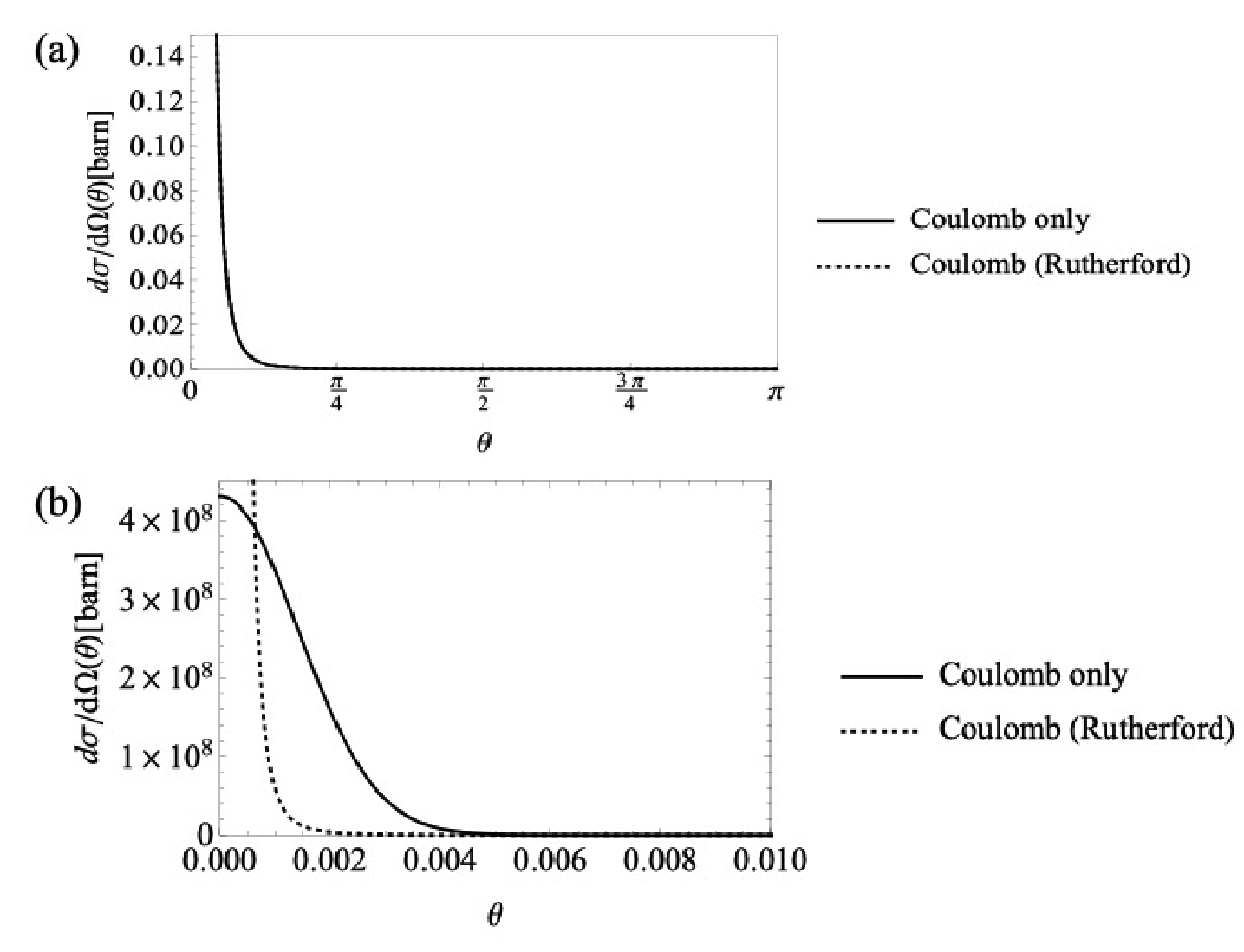}
\par\end{centering}
\caption{\label{fig:Differential-cross-section-2}Differential cross section
for the Coulomb interaction only, with model parameters.}

\end{figure}

In fig. \ref{fig:Differential-cross-section-2}, we plot (on the same
scale as that of fig. (\ref{fig:Differential-cross-section-1}) the
cross section for the Coulomb interaction alone, compared to the Rutherford
Coulomb result (\citet{Messiah1961,Rutherford1911}),
\[
\frac{d\sigma}{d\Omega}\vert_{\mathrm{Rutherford}}[\mathrm{barn}]=\frac{389}{p[\mathrm{MeV}]^{2}}\frac{\eta^{C2}}{4\sin^{4}(\frac{\theta}{2})}.
\]
No separation is used here. The Coulomb cross section does not generally
separate into a narrow forward peak and a finite contribution around
$\theta=0,$ as seen for shorter-range potentials, as we saw in \citet{Hoffmann2017a}.
This case is an exception, in the regime of low interaction strength,
where the cross section is dominated by the essentially free contribution
close to the origin.

\begin{figure}
\begin{centering}
\includegraphics[width=17cm]{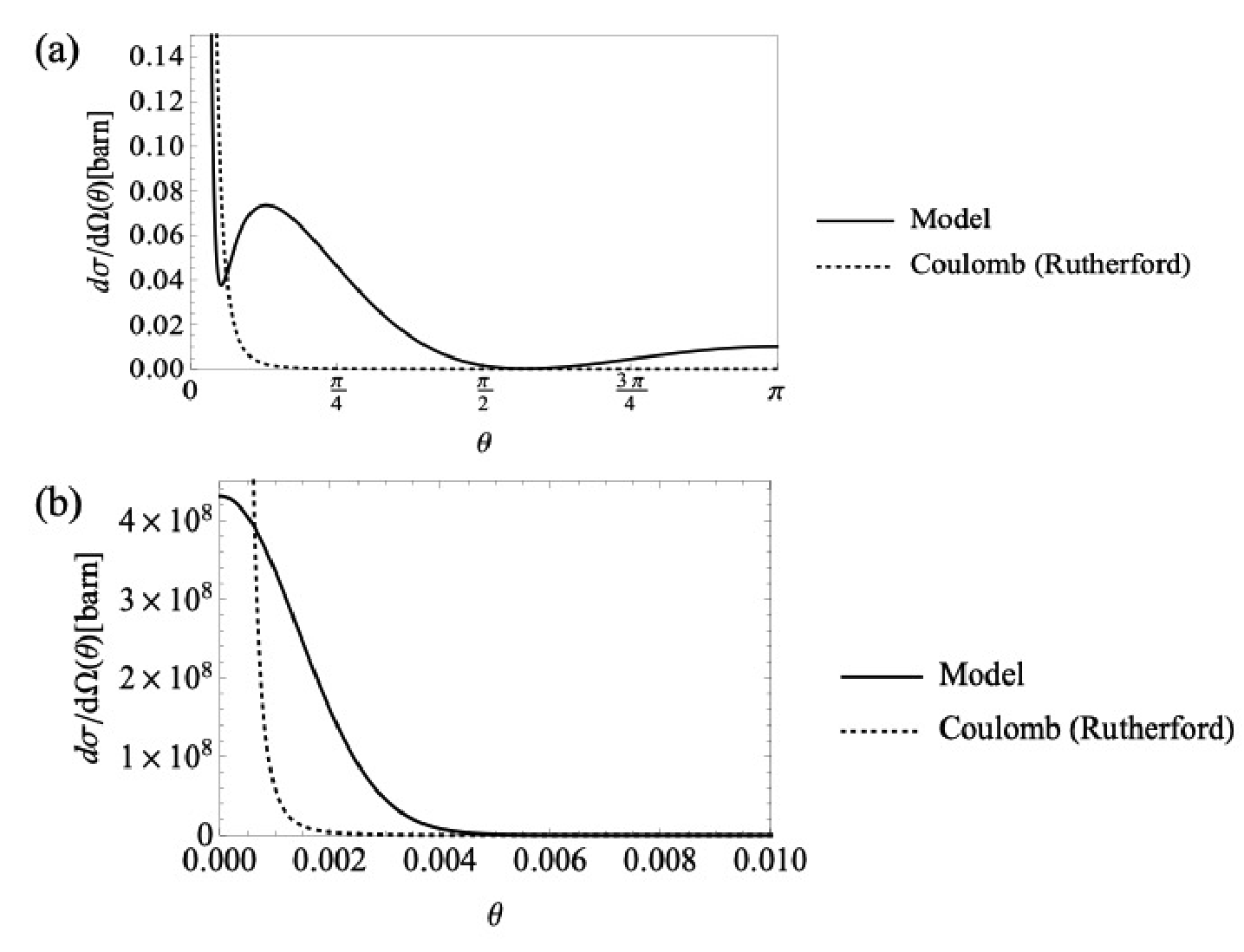}
\par\end{centering}
\caption{\label{fig:Differential-cross-section}Differential cross section
in barns as a function of angle, compared to the Rutherford result.}

\end{figure}

Results for the Coulomb plus nuclear model are show in fig. \ref{fig:Differential-cross-section}.
We see that the model cross section is significantly larger than the
Rutherford cross section at large angles. It is approximately the
incoherent sum of the nuclear and Coulomb parts at angles greater
than $\pi/4.$ At lower angles, the difference is larger, so an interference
term must be contributing significantly. We show the cross section
at low angles, where the wavepacket treatment guarantees finiteness.
The Rutherford result, derived classically, does not include that
constraint, but pure Coulomb cross sections with the wavepacket treatment
were also found to be finite at $\theta=0$ in \citet{Hoffmann2017a},
for a range of different parameters.

Votta \textit{et al.} (\citet{Votta1974}) measured the differential
cross section for protons on $\phantom{\vert}_{2}^{4}\mathrm{He}$
at an incident kinetic energy of $E_{V}=85\,\mathrm{MeV}$ ($p_{V}=408\,\mathrm{MeV}$).
The peak seen in figure \ref{fig:Differential-cross-section} is at
about $0.07\,\mathrm{b}.$ If we simply adjust this for the difference
in momentum by
\[
\left(\frac{p}{p_{V}}\right)^{2}0.07\,\mathrm{b}=\left(\frac{237}{408}\right)^{2}0.07\,\mathrm{b}=24\,\mathrm{mb},
\]
we find a value well within the significant region of their measurements.
It is remarkable that such a crude representation of the nuclear potential
leads to results of the correct order of magnitude.

\section{\label{sec:Conclusions}Conclusions}

We have constructed a perturbation theory to treat interactions that
can include the Coulomb interaction, such as is encountered in nuclear
physics. The Coulomb part is not treated perturbatively; the exact
solutions are employed.

The first task was to extend the perturbation theory for non-Coulomb
interactions from the case where the angular momentum quantum number
was $l=0$ (the only case considered in \citet{Hoffmann2021a}) to
general nonnegative integers, $l.$ The method was tested on a system
with an exact solution, up to second order in the perturbation, for
$l=0$ and $l=5$. As expected, an $\mathcal{O}(\eta^{3})$ error
was found in both cases, where $\eta$ is the dimensionless coupling
strength (eq. (\ref{eq:2.16.5})). Analysis of the second-order terms
in the series shows that we expect finiteness for general potentials,
provided their singularity at the origin is less than $1/r^{2}$ and
that they fall off asymptotically with $r$ faster than $1/r.$

Subsequently, the method for the sum of a Coulomb potential and a
perturbing potential was developed, and found to have similarities
with the non-Coulomb method. A model system was simulated, involving
a Coulomb interaction and a spherical well, the latter being a primitive
representation of a nuclear interaction. The differential cross section
for scattering from this potential was calculated, using the wavepacket
formalism of \citet{Hoffmann2017a}. The physically realistic result
differs from the incoherent sum of Coulomb and nuclear cross sections,
as expected. The magnitudes of the differential cross sections were
realistic.

Again, the second-order terms in the perturbation series were finite.
Analysis shows that they would continue to be finite for general perturbing
potentials in the class of sufficiently well-behaved potentials that
we have defined. We conjecture that the terms would remain finite
at third and higher order. This is the central result of this paper:
a finite perturbation theory that can include the Coulomb potential.
In contrast, applying perturbation theory directly to the Coulomb
potential gives infinities.

\appendix

\section{Evaluation of integrals for perturbation about the free solutions}

In \citet{Hoffmann2021a} we derived results only for the case $l=0.$
Here we evaluate integrals that we will need for general $l=0,1,2,\dots$

The asymptotic form of the free spherical waves is (\citet{Messiah1961})
\begin{equation}
y_{l}^{(0)}(r,k)\rightarrow\sqrt{\frac{2}{\pi}}\,\sin(kr-l\frac{\pi}{2})\label{eq:a.1}
\end{equation}
as $kr\rightarrow\infty.$ We will find two types of principal part
integrals that we will need to evaluate in the same asymptotic limit:
\begin{align}
I_{1} & =P\int_{0}^{\infty}dk\,\sqrt{\frac{2}{\pi}}\,\sin(kr-l\frac{\pi}{2})\frac{f(k)}{k^{2}-p^{2}},\nonumber \\
I_{2} & =P\int_{0}^{\infty}dk\,\sqrt{\frac{2}{\pi}}\,\cos(kr-l\frac{\pi}{2})\frac{f(k)}{k^{2}-p^{2}}.\label{eq:a.2}
\end{align}
Here $f(k)$ is analytic on the integration region and is such that
the integrals always converge.

We separate these into integrals on $k\in[0,2p]$ and integrals on
$k\in[2p,\infty).$ For the finite integration region, we define $k=p(1+x)$
and $\kappa=pr,$ and use
\begin{align}
\sin(\kappa-l\frac{\pi}{2}+\kappa x) & =\sin(\kappa-l\frac{\pi}{2})\cos(\kappa x)+\cos(\kappa-l\frac{\pi}{2})\sin(\kappa x),\nonumber \\
\cos(\kappa-l\frac{\pi}{2}+\kappa x) & =\cos(\kappa-l\frac{\pi}{2})\cos(\kappa x)-\sin(\kappa-l\frac{\pi}{2})\sin(\kappa x).\label{eq:a.3}
\end{align}
Also for the integrals on $x\in[-1,1],$ we separate the integrands
into parts even and odd in $x.$ The latter will vanish under principal
part integration, including when the integrand has a $1/x$ singularity.
The remaining integrand will be free from singularities and standard
integration can be used. The two integrals on $k\in[2p,\infty)$ have
no singularities on that region, so reduce to standard integrals.
It will then suffice to consider
\begin{align}
I_{\mathrm{0}} & =\int_{-1}^{1}dx\,\frac{\sin(\kappa x)}{x}f_{+}(x),\nonumber \\
I_{s,F} & =\int_{-1}^{1}dx\,\sin(\kappa x)f_{+}(x),\quad I_{c,F}=\int_{-1}^{1}dx\,\cos(\kappa x)f_{+}(x),\nonumber \\
I_{s,I} & =\int_{2}^{\infty}dz\,\sin(pz)\,g(z),\quad I_{c,I}=\int_{2}^{\infty}dz\,\cos(pz)\,g(z)\label{eq:a.4}
\end{align}
where $f_{+}(x)$ is even in $x$ and satisfies our regularity and
convergence requirements. For the infinite integration region, we
used $k=pz.$ Here $g(z)$ must decrease at least as fast as $1/z$
for convergence.

In the first of these integrals, the functions
\begin{equation}
\Delta(x,\kappa)=\frac{\sin(\kappa x)}{x}\label{eq:a.5}
\end{equation}
have peak value $\Delta(0,\kappa)=\kappa\gg1$ and fall off like $1/x,$
with rapid oscillations, over a width of order $1/\kappa.$ Thus they
are approximations to a delta function, the approximation improving
as $\kappa\rightarrow\infty.$ They are normalized to
\begin{equation}
\int_{-1}^{1}dx\,\frac{\sin(\kappa x)}{x}=\pi+\mathcal{O}(\frac{1}{\kappa}).\label{eq:a.6}
\end{equation}
So we expect
\begin{equation}
\Delta(x,\kappa)\rightarrow\pi\,\delta(x)\label{eq:a.7}
\end{equation}
as $\kappa\rightarrow\infty.$

Hence we expect $I_{\mathrm{0}}=\pi f_{+}(0).$ Integration by parts
applied to the remaining integrals shows that they will vanish like
order $1/\kappa.$ We verified the $I_{\mathrm{0}}$ result numerically,
using polynomials in $x^{2}$ for $f_{+}(x).$

We find that together these results give the asymptotic forms
\begin{equation}
P\int_{0}^{\infty}dk\,\sqrt{\frac{2}{\pi}}\,\sin(kr-l\frac{\pi}{2})\frac{f(k)}{k^{2}-p^{2}}\rightarrow\{\frac{\pi}{2p}f(p)\}\sqrt{\frac{2}{\pi}}\,\cos(pr-l\frac{\pi}{2})\label{eq:a.8}
\end{equation}
and
\begin{equation}
P\int_{0}^{\infty}dk\,\sqrt{\frac{2}{\pi}}\,\cos(kr-l\frac{\pi}{2})\frac{f(k)}{k^{2}-p^{2}}\rightarrow\{-\frac{\pi}{2p}f(p)\}\sqrt{\frac{2}{\pi}}\,\sin(pr-l\frac{\pi}{2}).\label{eq:a.9}
\end{equation}

\section{Evaluation of integrals for perturbation about the Coulomb solutions}

The asymptotic form of the Coulomb wavefunctions as $kr\rightarrow\infty$
is (\citet{Messiah1961})
\begin{equation}
y_{l}^{(C)}(r,k)=\sqrt{\frac{2}{\pi}}\,\sin(kr-\eta(k)\ln(2kr)-l\frac{\pi}{2}+\sigma_{l}[\eta(k)]),\label{eq:b.1}
\end{equation}
with
\begin{equation}
\eta(k)=\frac{Z_{t}Z_{p}\,\alpha}{k/m}.\label{eq:b.2}
\end{equation}
This asymptotic form differs from the free case by the presence of
the logarithmic phase, $-\eta(k)\ln(2kr),$ and the Coulomb phase
shifts, $\sigma_{l}[\eta(k)].$ It is not obvious that we will be
able to obtain results similar to those just derived for the free
case.

The phase
\begin{equation}
\varphi_{l}(r,k)=kr-\eta(k)\ln(2kr)-l\frac{\pi}{2}+\sigma_{l}[\eta(k)]\label{eq:b.3}
\end{equation}
has derivative
\begin{equation}
\frac{\partial\varphi_{l}(r,k)}{\partial k}=r+\frac{Z_{t}Z_{p}\,\alpha m}{k^{2}}\{\ln(\frac{2kr}{e})-\mathrm{Re}(\psi(l+1+i\eta(k)))\}.\label{eq:b.4}
\end{equation}
We used (\citet{Messiah1961})
\begin{equation}
e^{i2\sigma_{l}(k)}=\frac{\Gamma(l+1+i\eta(k))}{\Gamma(l+1-i\eta(k))}\label{eq:b.5}
\end{equation}
and
\begin{equation}
\psi(z)=\frac{1}{\Gamma(z)}\frac{d\Gamma(z)}{dz},\label{eq:b.6}
\end{equation}
the psi or Polygamma function (\citet{Gradsteyn1980}, their section
8.36).

This first derivative will generally be large because of the presence
of $r,$ giving a rapidly oscillating sine function, but the phase
has a stationary point where
\begin{equation}
1+\frac{Z_{t}Z_{p}\,\alpha}{v^{2}\rho}\{\ln(\frac{2v\rho}{e})-\mathrm{Re}(\psi(l+1+i\frac{Z_{t}Z_{p}\,\alpha}{v}))\}=0,\label{eq:b.7}
\end{equation}
where
\begin{equation}
v=\frac{k}{m},\quad\rho=mr,\label{eq:b.8}
\end{equation}
both dimensionless. In the neighbourhood of the stationary point,
the phase varies slowly and we will find an undesired finite contribution
to the integrals we will consider (those in parallel to eqs. (\ref{eq:a.4})).

We find that the root of eq. (\ref{eq:b.7}), $v_{0}(\rho),$ approaches
zero from above as $\rho\rightarrow\infty.$ The state vector with
$k=0$ is unphysical in our theory. It has a wavefunction independent
of $r$ to which we cannot apply continuum normalization. Hence we
argue that the stationary point should have no physical effect as
$\rho\rightarrow\infty.$ Numerically, it is a simple matter to remove
its contribution for finite $r$. We find a small region, $v\in[0,\epsilon(\rho)],$
containing the stationary point, with, say $\epsilon(\rho)=2v_{0}(\rho),$
and remove that region from the integrals. In terms of $x,$ this
region is $x\in[-1,-1+\frac{2\epsilon(\rho)}{p/m}].$

We want to evaluate integrals similar to those in eq. (\ref{eq:a.4}).
We define
\begin{equation}
\Delta\varphi_{l}(\rho,v,x)=\varphi_{l}(r,p(1+x))-\varphi_{l}(r,p)\label{eq:b.9}
\end{equation}
so that the Taylor series around $x=0$ is
\begin{equation}
\Delta\varphi_{l}(\rho,v,x)=px\frac{\partial\varphi_{l}(r,k)}{\partial k}|_{k=p}+\mathcal{O}(x^{2}).\label{eq:b.10}
\end{equation}
This gives integrals that are linear combinations of $\sin(\varphi_{l}(r,p))$
and $\cos(\varphi_{l}(r,p)).$ So we need to evalute the remaining
integrals of the form
\begin{align}
I_{\mathrm{0}}^{(C)} & =\int_{-1}^{1}dx\,\frac{\sin(\Delta\varphi_{l}(\rho,v,x))}{x}f_{+}(x),\nonumber \\
I_{s,F}^{(C)} & =\int_{-1}^{1}dx\,\sin(\Delta\varphi_{l}(\rho,v,x))f_{+}(x),\quad I_{c,F}^{(C)}=\int_{-1}^{1}dx\,\cos(\Delta\varphi_{l}(\rho,v,x))f_{+}(x),\nonumber \\
I_{s,I}^{(C)} & =\int_{2}^{\infty}dz\,\sin(\Delta\varphi_{l}(\rho,v,x))\,g(z),\quad I_{c,I}^{(C)}=\int_{2}^{\infty}dz\,\cos(\Delta\varphi_{l}(\rho,v,x))\,g(z),\label{eq:b.11}
\end{align}
with the above conditions on $f_{+}(x)$ and $g(z).$ We find that
the last four integrals all vanish like $1/v\rho=1/pr$ as $pr\rightarrow\infty.$

Since the dominant term in the phase derivative gives
\begin{equation}
px\frac{\partial\varphi_{l}(r,k)}{\partial k}\vert_{k=p}\sim prx,\label{eq:b.12}
\end{equation}
we define
\begin{equation}
\Delta\varphi_{l}(\rho,v,x)=prx+\delta\varphi_{l}(\rho,v,x),\label{eq:b.13}
\end{equation}
giving
\begin{equation}
I_{0}^{(C)}=\int_{-1}^{1}dx\,\frac{\sin(v\rho x)}{x}F_{+}(x)+\int_{-1}^{1}dx\,\cos(v\rho x)G_{+}(x),\label{eq:b.14}
\end{equation}
with
\begin{align}
F_{+}(x) & =[\cos(\delta\varphi_{l}(v,\rho,x))]_{+}f_{+}(x),\nonumber \\
G_{+}(x) & =\frac{[\sin(\delta\varphi_{l}(v,\rho,x))]_{-}}{x}f_{+}(x).\label{eq:b.15}
\end{align}

We apply integration by parts to both terms, giving
\begin{multline}
I_{0}^{(C)}=\{\pi f_{+}(0)+2\mathrm{si}(v\rho)F_{+}(1)\\
\quad-2\int_{0}^{1}dx\,\mathrm{si}(v\rho x)\,F_{+}^{\prime}(x)\}+\{\frac{2G_{1}(1)\sin(v\rho x)}{v\rho}-\frac{1}{v\rho}\int_{0}^{1}dx\,\sin(v\rho x)\,G_{+}^{\prime}(x)\},\label{eq:b.16}
\end{multline}
where (\citet{Gradsteyn1980}, their section 8.23)
\begin{equation}
\mathrm{si}(z)=-\int_{z}^{\infty}dt\,\frac{\sin t}{t}.\label{eq:b.17}
\end{equation}
We note $\mathrm{si}(v\rho)\sim\mathcal{O}(1/v\rho)$ as $v\rho\rightarrow\infty.$

We encountered difficulties trying to numerically integrate the integrals
in eq. (\ref{eq:b.11}) directly. The problem is integrands with rapidly
oscillatory behaviour. This analytic treatment leads to two remainders,
\begin{equation}
R_{1}=-2\int_{0}^{1}dx\,\mathrm{si}(v\rho x)\,F_{+}^{\prime}(x)\quad\mathrm{and}\quad R_{2}=-\frac{1}{v\rho}\int_{0}^{1}dx\,\sin(v\rho x)\,G_{+}^{\prime}(x),\label{eq:b.18}
\end{equation}
both of which could be numerically integrated. Both were found to
vanish as $v\rho\rightarrow\infty.$ So we find
\begin{equation}
\lim_{pr\rightarrow\infty}I_{\mathrm{sinc}}^{(C)}=\pi f_{+}(0).\label{eq:b.19}
\end{equation}

Using these results, we find

\begin{multline}
P\int_{0}^{\infty}dk\,\sqrt{\frac{2}{\pi}}\,\sin(kr-\eta(k)\ln(2kr)-l\frac{\pi}{2}+\sigma_{l}[\eta(k)])\frac{f(k)}{k^{2}-p^{2}}\rightarrow\\
\{\frac{\pi}{2p}f(p)\}\sqrt{\frac{2}{\pi}}\,\cos(pr-\eta(p)\ln(2pr)-l\frac{\pi}{2}+\sigma_{l}[\eta(p)])\label{eq:b.20}
\end{multline}
and
\begin{multline}
P\int_{0}^{\infty}dk\,\sqrt{\frac{2}{\pi}}\,\cos(kr-\eta(k)\ln(2kr)-l\frac{\pi}{2}+\sigma_{l}[\eta(k)])\frac{f(k)}{k^{2}-p^{2}}\rightarrow\\
\{-\frac{\pi}{2p}f(p)\}\sqrt{\frac{2}{\pi}}\,\sin(pr-\eta(p)\ln(2pr)-l\frac{\pi}{2}+\sigma_{l}[\eta(p)]),\label{eq:b.21}
\end{multline}
where $f(k)$ is regular on the integration region and such that the
integrals always converge.\medskip{}

Data sharing is not applicable to this article as no new data were
created or analyzed in this study. 

\bibliographystyle{apsrev4-1}

\begin{thebibliography}{22}%
\makeatletter
\providecommand \@ifxundefined [1]{%
 \@ifx{#1\undefined}
}%
\providecommand \@ifnum [1]{%
 \ifnum #1\expandafter \@firstoftwo
 \else \expandafter \@secondoftwo
 \fi
}%
\providecommand \@ifx [1]{%
 \ifx #1\expandafter \@firstoftwo
 \else \expandafter \@secondoftwo
 \fi
}%
\providecommand \natexlab [1]{#1}%
\providecommand \enquote  [1]{``#1''}%
\providecommand \bibnamefont  [1]{#1}%
\providecommand \bibfnamefont [1]{#1}%
\providecommand \citenamefont [1]{#1}%
\providecommand \href@noop [0]{\@secondoftwo}%
\providecommand \href [0]{\begingroup \@sanitize@url \@href}%
\providecommand \@href[1]{\@@startlink{#1}\@@href}%
\providecommand \@@href[1]{\endgroup#1\@@endlink}%
\providecommand \@sanitize@url [0]{\catcode `\\12\catcode `\$12\catcode
  `\&12\catcode `\#12\catcode `\^12\catcode `\_12\catcode `\%12\relax}%
\providecommand \@@startlink[1]{}%
\providecommand \@@endlink[0]{}%
\providecommand \url  [0]{\begingroup\@sanitize@url \@url }%
\providecommand \@url [1]{\endgroup\@href {#1}{\urlprefix }}%
\providecommand \urlprefix  [0]{URL }%
\providecommand \Eprint [0]{\href }%
\providecommand \doibase [0]{http://dx.doi.org/}%
\providecommand \selectlanguage [0]{\@gobble}%
\providecommand \bibinfo  [0]{\@secondoftwo}%
\providecommand \bibfield  [0]{\@secondoftwo}%
\providecommand \translation [1]{[#1]}%
\providecommand \BibitemOpen [0]{}%
\providecommand \bibitemStop [0]{}%
\providecommand \bibitemNoStop [0]{.\EOS\space}%
\providecommand \EOS [0]{\spacefactor3000\relax}%
\providecommand \BibitemShut  [1]{\csname bibitem#1\endcsname}%
\let\auto@bib@innerbib\@empty
\bibitem [{\citenamefont {Deltuva}\ \emph {et~al.}(2005)\citenamefont
  {Deltuva}, \citenamefont {Fonseca}, \citenamefont {Kievsky}, \citenamefont
  {Rosati}, \citenamefont {Sauer},\ and\ \citenamefont
  {Viviani}}]{Deltuva2005}%
  \BibitemOpen
  \bibfield  {author} {\bibinfo {author} {\bibfnamefont {A.}~\bibnamefont
  {Deltuva}}, \bibinfo {author} {\bibfnamefont {A.~C.}\ \bibnamefont
  {Fonseca}}, \bibinfo {author} {\bibfnamefont {A.}~\bibnamefont {Kievsky}},
  \bibinfo {author} {\bibfnamefont {S.}~\bibnamefont {Rosati}}, \bibinfo
  {author} {\bibfnamefont {P.~U.}\ \bibnamefont {Sauer}}, \ and\ \bibinfo
  {author} {\bibfnamefont {M.}~\bibnamefont {Viviani}},\ }\href {\doibase
  10.1103/PhysRevC.71.064003} {\bibfield  {journal} {\bibinfo  {journal} {Phys.
  Rev. C}\ }\textbf {\bibinfo {volume} {71}},\ \bibinfo {pages} {064003}
  (\bibinfo {year} {2005})}\BibitemShut {NoStop}%
\bibitem [{\citenamefont {Durand}\ and\ \citenamefont {Ha}(2020)}]{Durand2020}%
  \BibitemOpen
  \bibfield  {author} {\bibinfo {author} {\bibfnamefont {L.}~\bibnamefont
  {Durand}}\ and\ \bibinfo {author} {\bibfnamefont {P.}~\bibnamefont {Ha}},\
  }\href {\doibase 10.1103/PhysRevD.102.036025} {\bibfield  {journal} {\bibinfo
   {journal} {Phys. Rev. D}\ }\textbf {\bibinfo {volume} {102}},\ \bibinfo
  {pages} {036025} (\bibinfo {year} {2020})}\BibitemShut {NoStop}%
\bibitem [{\citenamefont {Franco}(1973)}]{Franco1973}%
  \BibitemOpen
  \bibfield  {author} {\bibinfo {author} {\bibfnamefont {V.}~\bibnamefont
  {Franco}},\ }\href {\doibase 10.1103/PhysRevD.7.215} {\bibfield  {journal}
  {\bibinfo  {journal} {Phys. Rev. D}\ }\textbf {\bibinfo {volume} {7}},\
  \bibinfo {pages} {215} (\bibinfo {year} {1973})}\BibitemShut {NoStop}%
\bibitem [{\citenamefont {Islam}(1967)}]{Islam1967}%
  \BibitemOpen
  \bibfield  {author} {\bibinfo {author} {\bibfnamefont {M.~M.}\ \bibnamefont
  {Islam}},\ }\href {\doibase 10.1103/PhysRev.162.1426} {\bibfield  {journal}
  {\bibinfo  {journal} {Phys. Rev.}\ }\textbf {\bibinfo {volume} {162}},\
  \bibinfo {pages} {1426} (\bibinfo {year} {1967})}\BibitemShut {NoStop}%
\bibitem [{\citenamefont {Petrov}(2018)}]{Petrov2018}%
  \BibitemOpen
  \bibfield  {author} {\bibinfo {author} {\bibfnamefont {V.~A.}\ \bibnamefont
  {Petrov}},\ }\href {\doibase 10.1140/epjc/s10052-018-5716-1} {\bibfield
  {journal} {\bibinfo  {journal} {Eur. Phys. J. C.}\ }\textbf {\bibinfo
  {volume} {78}},\ \bibinfo {pages} {221} (\bibinfo {year} {2018})}\BibitemShut
  {NoStop}%
\bibitem [{\citenamefont {West}\ and\ \citenamefont {Yennie}(1968)}]{West1968}%
  \BibitemOpen
  \bibfield  {author} {\bibinfo {author} {\bibfnamefont {G.~B.}\ \bibnamefont
  {West}}\ and\ \bibinfo {author} {\bibfnamefont {D.~R.}\ \bibnamefont
  {Yennie}},\ }\href {\doibase 10.1103/PhysRev.172.1413} {\bibfield  {journal}
  {\bibinfo  {journal} {Phys. Rev.}\ }\textbf {\bibinfo {volume} {172}},\
  \bibinfo {pages} {1413} (\bibinfo {year} {1968})}\BibitemShut {NoStop}%
\bibitem [{\citenamefont {Hoffmann}(2017)}]{Hoffmann2017a}%
  \BibitemOpen
  \bibfield  {author} {\bibinfo {author} {\bibfnamefont {S.~E.}\ \bibnamefont
  {Hoffmann}},\ }\href@noop {} {\bibfield  {journal} {\bibinfo  {journal} {J.
  Phys. B: At. Mol. Opt. Phys.}\ }\textbf {\bibinfo {volume} {50}},\ \bibinfo
  {pages} {215302} (\bibinfo {year} {2017})}\BibitemShut {NoStop}%
\bibitem [{\citenamefont {Dalitz}(1951)}]{Dalitz1951}%
  \BibitemOpen
  \bibfield  {author} {\bibinfo {author} {\bibfnamefont {R.~H.}\ \bibnamefont
  {Dalitz}},\ }\href@noop {} {\bibfield  {journal} {\bibinfo  {journal} {Proc.
  Roy. Soc. Lond. A: Math., Phys. and Eng. Sci.}\ }\textbf {\bibinfo {volume}
  {206}},\ \bibinfo {pages} {509} (\bibinfo {year} {1951})}\BibitemShut
  {NoStop}%
\bibitem [{\citenamefont {Hoffmann}(2021)}]{Hoffmann2021a}%
  \BibitemOpen
  \bibfield  {author} {\bibinfo {author} {\bibfnamefont {S.~E.}\ \bibnamefont
  {Hoffmann}},\ }\href {\doibase 10.1063/5.0023630} {\bibfield  {journal}
  {\bibinfo  {journal} {J. Math. Phys.}\ }\textbf {\bibinfo {volume} {62}},\
  \bibinfo {pages} {032105} (\bibinfo {year} {2021})}\BibitemShut {NoStop}%
\bibitem [{\citenamefont {Vigo-Aguiar}\ and\ \citenamefont
  {Simos}(2005)}]{Vigo-Aguiar2005}%
  \BibitemOpen
  \bibfield  {author} {\bibinfo {author} {\bibfnamefont {J.}~\bibnamefont
  {Vigo-Aguiar}}\ and\ \bibinfo {author} {\bibfnamefont {T.~E.}\ \bibnamefont
  {Simos}},\ }\href {\doibase https://doi.org/10.1002/qua.20495} {\bibfield
  {journal} {\bibinfo  {journal} {Int. J. Quantum Chem.}\ }\textbf {\bibinfo
  {volume} {103}},\ \bibinfo {pages} {278} (\bibinfo {year}
  {2005})}\BibitemShut {NoStop}%
\bibitem [{\citenamefont {Simos}\ and\ \citenamefont
  {Williams}(1997)}]{Simos1997}%
  \BibitemOpen
  \bibfield  {author} {\bibinfo {author} {\bibfnamefont {T.}~\bibnamefont
  {Simos}}\ and\ \bibinfo {author} {\bibfnamefont {P.}~\bibnamefont
  {Williams}},\ }\href {\doibase https://doi.org/10.1016/S0097-8485(96)00024-1}
  {\bibfield  {journal} {\bibinfo  {journal} {Comput. Chem.(Oxford)}\ }\textbf
  {\bibinfo {volume} {21}},\ \bibinfo {pages} {175} (\bibinfo {year}
  {1997})}\BibitemShut {NoStop}%
\bibitem [{\citenamefont {Messiah}(1961)}]{Messiah1961}%
  \BibitemOpen
  \bibfield  {author} {\bibinfo {author} {\bibfnamefont {A.}~\bibnamefont
  {Messiah}},\ }\href@noop {} {\emph {\bibinfo {title} {Quantum Mechanics}}},\
  Vol.\ \bibinfo {volume} {1 and 2}\ (\bibinfo  {publisher} {North-Holland,
  Amsterdam and John Wiley and Sons, N.Y.},\ \bibinfo {year}
  {1961})\BibitemShut {NoStop}%
\bibitem [{\citenamefont {Kramers}\ and\ \citenamefont
  {Ittman}(1929)}]{Kramers1929}%
  \BibitemOpen
  \bibfield  {author} {\bibinfo {author} {\bibfnamefont {H.~A.}\ \bibnamefont
  {Kramers}}\ and\ \bibinfo {author} {\bibfnamefont {G.~P.}\ \bibnamefont
  {Ittman}},\ }\href@noop {} {\bibfield  {journal} {\bibinfo  {journal} {Z.
  Physik}\ }\textbf {\bibinfo {volume} {58}},\ \bibinfo {pages} {225} (\bibinfo
  {year} {1929})}\BibitemShut {NoStop}%
\bibitem [{\citenamefont {Bethe}(1958)}]{Bethe1958}%
  \BibitemOpen
  \bibfield  {author} {\bibinfo {author} {\bibfnamefont {H.}~\bibnamefont
  {Bethe}},\ }\href {\doibase https://doi.org/10.1016/0003-4916(58)90017-4}
  {\bibfield  {journal} {\bibinfo  {journal} {Ann. Phys. (N.Y.)}\ }\textbf
  {\bibinfo {volume} {3}},\ \bibinfo {pages} {190} (\bibinfo {year}
  {1958})}\BibitemShut {NoStop}%
\bibitem [{\citenamefont {Blanes}\ \emph {et~al.}(2009)\citenamefont {Blanes},
  \citenamefont {Casas}, \citenamefont {Oteo},\ and\ \citenamefont
  {Ros}}]{Blanes2009}%
  \BibitemOpen
  \bibfield  {author} {\bibinfo {author} {\bibfnamefont {S.}~\bibnamefont
  {Blanes}}, \bibinfo {author} {\bibfnamefont {F.}~\bibnamefont {Casas}},
  \bibinfo {author} {\bibfnamefont {J.}~\bibnamefont {Oteo}}, \ and\ \bibinfo
  {author} {\bibfnamefont {J.}~\bibnamefont {Ros}},\ }\href {\doibase
  https://doi.org/10.1016/j.physrep.2008.11.001} {\bibfield  {journal}
  {\bibinfo  {journal} {Phys. Rep.}\ }\textbf {\bibinfo {volume} {470}},\
  \bibinfo {pages} {151} (\bibinfo {year} {2009})}\BibitemShut {NoStop}%
\bibitem [{\citenamefont {Blanes}\ \emph {et~al.}(2010)\citenamefont {Blanes},
  \citenamefont {Casas}, \citenamefont {Oteo},\ and\ \citenamefont
  {Ros}}]{Blanes2010}%
  \BibitemOpen
  \bibfield  {author} {\bibinfo {author} {\bibfnamefont {S.}~\bibnamefont
  {Blanes}}, \bibinfo {author} {\bibfnamefont {F.}~\bibnamefont {Casas}},
  \bibinfo {author} {\bibfnamefont {J.~A.}\ \bibnamefont {Oteo}}, \ and\
  \bibinfo {author} {\bibfnamefont {J.}~\bibnamefont {Ros}},\ }\href {\doibase
  10.1088/0143-0807/31/4/020} {\bibfield  {journal} {\bibinfo  {journal}
  {European Journal of Physics}\ }\textbf {\bibinfo {volume} {31}},\ \bibinfo
  {pages} {907} (\bibinfo {year} {2010})}\BibitemShut {NoStop}%
\bibitem [{\citenamefont {Casas}(2007)}]{Casas2007}%
  \BibitemOpen
  \bibfield  {author} {\bibinfo {author} {\bibfnamefont {F.}~\bibnamefont
  {Casas}},\ }\href {\doibase 10.1088/1751-8113/40/50/006} {\bibfield
  {journal} {\bibinfo  {journal} {J. Phys. A: Math. Theor.}\ }\textbf {\bibinfo
  {volume} {40}},\ \bibinfo {pages} {15001} (\bibinfo {year}
  {2007})}\BibitemShut {NoStop}%
\bibitem [{\citenamefont {Abramowitz}\ and\ \citenamefont
  {Stegun}(1972)}]{Abramowitz1972}%
  \BibitemOpen
  \bibfield  {author} {\bibinfo {author} {\bibfnamefont {M.}~\bibnamefont
  {Abramowitz}}\ and\ \bibinfo {author} {\bibfnamefont {I.~A.}\ \bibnamefont
  {Stegun}},\ }\href@noop {} {\emph {\bibinfo {title} {Handbook of Mathematical
  Functions}}},\ \bibinfo {edition} {9th}\ ed.\ (\bibinfo  {publisher} {Dover,
  N. Y.},\ \bibinfo {year} {1972})\BibitemShut {NoStop}%
\bibitem [{\citenamefont {Gradsteyn}\ and\ \citenamefont
  {Ryzhik}(1980)}]{Gradsteyn1980}%
  \BibitemOpen
  \bibfield  {author} {\bibinfo {author} {\bibfnamefont {I.~S.}\ \bibnamefont
  {Gradsteyn}}\ and\ \bibinfo {author} {\bibfnamefont {I.~M.}\ \bibnamefont
  {Ryzhik}},\ }\href@noop {} {\emph {\bibinfo {title} {Tables of Integrals,
  Series and Products}}},\ \bibinfo {edition} {corrected and enlarged}\ ed.\
  (\bibinfo  {publisher} {Academic Press, Inc., San Diego, CA},\ \bibinfo
  {year} {1980})\BibitemShut {NoStop}%
\bibitem [{\citenamefont {Czachor}\ and\ \citenamefont {P{\k
  e}czkowski}(2011)}]{Czachor2011}%
  \BibitemOpen
  \bibfield  {author} {\bibinfo {author} {\bibfnamefont {A.}~\bibnamefont
  {Czachor}}\ and\ \bibinfo {author} {\bibfnamefont {P.}~\bibnamefont {P{\k
  e}czkowski}},\ }\href@noop {} {\bibfield  {journal} {\bibinfo  {journal}
  {Phys. Part. Nucl. Lett.}\ }\textbf {\bibinfo {volume} {8}} (\bibinfo {year}
  {2011})}\BibitemShut {NoStop}%
\bibitem [{\citenamefont {Rutherford}(1911)}]{Rutherford1911}%
  \BibitemOpen
  \bibfield  {author} {\bibinfo {author} {\bibfnamefont {E.}~\bibnamefont
  {Rutherford}},\ }\href@noop {} {\bibfield  {journal} {\bibinfo  {journal}
  {Phil. Mag.}\ }\textbf {\bibinfo {volume} {Series 6, vol. 21}},\ \bibinfo
  {pages} {669} (\bibinfo {year} {1911})}\BibitemShut {NoStop}%
\bibitem [{\citenamefont {Votta}\ \emph {et~al.}(1974)\citenamefont {Votta},
  \citenamefont {Roos}, \citenamefont {Chant},\ and\ \citenamefont
  {Woody}}]{Votta1974}%
  \BibitemOpen
  \bibfield  {author} {\bibinfo {author} {\bibfnamefont {L.~G.}\ \bibnamefont
  {Votta}}, \bibinfo {author} {\bibfnamefont {P.~G.}\ \bibnamefont {Roos}},
  \bibinfo {author} {\bibfnamefont {N.~S.}\ \bibnamefont {Chant}}, \ and\
  \bibinfo {author} {\bibfnamefont {R.}~\bibnamefont {Woody}},\ }\href
  {\doibase 10.1103/PhysRevC.10.520} {\bibfield  {journal} {\bibinfo  {journal}
  {Phys. Rev. C}\ }\textbf {\bibinfo {volume} {10}},\ \bibinfo {pages} {520}
  (\bibinfo {year} {1974})}\BibitemShut {NoStop}%
\end{thebibliography}

\end{document}